\documentclass[12pt]{iopart}
\usepackage{iopams}
\usepackage{setstack}
\usepackage{graphicx}

\newcommand{\half}{\textstyle{\frac{1}{2}}}

\newcommand{\cP}{{\cal P}}
\newcommand{\cT}{{\cal T}}

\begin{document}

\title[Classical Particle in a Complex Elliptic Potential] {Classical Particle
in a Complex Elliptic Potential}

\author[Bender, Hook, and Kooner]{Carl~M~Bender${}^\ast$, Daniel~W~Hook${}^\dag$
and Karta Singh Kooner${}^\dag$}

\address{${}^\ast$Department of Physics, Washington University, St. Louis, MO
63130, USA \\{\footnotesize{\tt email: cmb@wustl.edu}}}

\address{${}^\dag$Theoretical Physics, Imperial College, London SW7 2AZ,
UK\\ {\footnotesize{\tt email: d.hook@imperial.ac.uk}}}

\date{today}

\begin{abstract}
This paper reports a numerical study of complex classical trajectories of a
particle in an elliptic potential. This study of doubly-periodic potentials is a
natural sequel to earlier work on complex classical trajectories in
trigonometric potentials. For elliptic potentials there is a two-dimensional
array of identical cells in the complex plane, and each cell contains a pair
of turning points. The particle can travel both horizontally and vertically as
it visits these cells, and sometimes the particle is captured temporarily by a
pair of turning points. If the particle's energy lies in a conduction band, the
particle drifts through the lattice of cells and is never captured by the same
pair of turning points more than once. However, if the energy of the particle is
not in a conduction band, the particle can return to previously visited cells.
\end{abstract}


\submitto{\JPA}

\section{Introduction}
\label{s1}

The past decade has seen much research activity devoted to extending quantum
mechanics into the complex domain \cite{S1}. It has been shown that Dirac
Hermiticity may be generalized to include complex non-Dirac-Hermitian
Hamiltonians that are $\cP\cT$ symmetric. (A Hamiltonian is {\it
Dirac-Hermitian} if $H=H^\dag$, where $\dag$ represents combined complex
conjugation and matrix transposition. A Hamiltonian is $\cP\cT$ symmetric if it
is invariant under combined spatial reflection $\cP$ and time reversal $\cT$.)
These complex Hamiltonians are physically acceptable because (i) their
eigenvalues are all real, and (ii) they generate unitary time evolution
\cite{S2,S3,S4}. Research on $\cP\cT$ quantum mechanics has revealed interesting
and surprising new phenomena, and in the past year some of these new phenomena
have actually been observed in laboratory experiments \cite{S5,S6,S7}.

Motivated by a desire to understand $\cP\cT$ quantum mechanics better, there has
been a significant body of research on complex classical mechanics. Conventional
classical mechanics is the study of the real solutions to Hamilton's equations
of motion. In the recently established field of {\it complex classical
mechanics}, we expand this study to include {\it all} solutions, complex as
well as real, to Hamilton's equations \cite{S8,S9}.

Complex classical mechanics has proved to be useful, in part because it provides
a heuristic explanation of the quantum $\cP\cT$ phase transition between a {\it
broken} $\cP\cT$-symmetric phase (where some of the quantum eigenvalues are real
and some are complex) and an {\it unbroken} $\cP\cT$-symmetric phase (where all
of the quantum eigenvalues are real). In the broken phase of $\cP\cT$ quantum
mechanics the classical trajectories are open, and in the unbroken phase the
classical trajectories are closed and periodic \cite{S10}. Bohr-Sommerfeld
quantization clarifies the association between complex energy and open classical
trajectories versus real energy and closed classical trajectories. This
association, which is discussed in detail in Ref.~\cite{S11}, arises because the
Bohr-Sommerfeld quantization condition contains a path integral along a closed
contour: $\oint_C dx\,p=\left(n+\half\right)\pi$. Evidently, this quantization
condition can only be applied if the classical orbits are closed.

Studies of complex classical trajectories have also led to studies of
probability densities in the complex plane, which in turn have given rise to the
formulation of a complex correspondence principle \cite{S12}. In addition,
studies of complex classical mechanics have proved to be useful in their own
right and may lead to a deeper understanding of the transition from nonchaotic
to chaotic behavior \cite{S13}.

One particularly interesting area of research concerns the generalization of
complex classical mechanics from real to complex energy. It can be argued that
because of the time-energy uncertainty principle in quantum mechanics, $\Delta t
\Delta E\gtrsim\hbar$, one cannot measure an energy with absolute precision
unless one has an infinite amount of time in which to perform the measurement.
We then suppose that this uncertainty in the value of the energy exists in
classical mechanics and further assume that this uncertainty $\Delta E$ may be
complex. The surprise is that generalizing complex classical mechanics from
real to complex energy produces a theory that exhibits many of the qualitative
features of quantum mechanics \cite{S11,S14,S15}.

The principal effect of extending the energy from real to complex values is that
particle trajectories $x(t)$, which are curves in the complex-$x$ plane, cease
to be closed. For example, while the particle trajectories for the classical
anharmonic-oscillator Hamiltonian $H=p^2+x^4$ are all closed when the energy $E$
is real, the trajectories are open and spiral outward to infinity when ${\rm
Im}\,E\neq0$ \cite{S11}. More interestingly, while the trajectories $x(t)$ for a
double-well anharmonic oscillator Hamiltonian $H=p^2-x^2+x^4$ are closed in the 
complex-$x$ plane, they are no longer closed when $E$ is complex. However, the
remarkable property of these trajectories is that they do not spiral outward to
infinity. Rather, they run back and forth endlessly from the vicinity of one
well to the vicinity of the other well in a fashion that is reminiscent of
quantum tunneling. To be specific, if a particle is initially in the classically
allowed region on the real axis in the left well, the particle spirals outward,
crosses the imaginary axis, and then spirals {\it inward} into the right well.
Then, the particle spirals outward again and visits the left well. Thus, the two
wells in the potential act like strange attractors. This oscillatory process
continues forever, but the trajectory never crosses itself.

This complex double-well classical system is different from the corresponding
quantum-mechanical system in that a trajectory cannot cross itself unless the
classical trajectory is periodic, and thus a classical particle on an open
trajectory can never revisit any point in the complex plane. However, many
features of quantum mechanics are mirrored in the properties of this classical
behavior. For example, as the imaginary part of the classical energy increases,
the characteristic ``tunneling'' time (the time required for the particle to
spiral inward and outward around a pair of turning points) decreases inversely,
just as one would expect of a quantum particle. As in the case of quantum
tunneling, the particle spends a long time in proximity to a given pair of
turning points before crossing the imaginary axis to the other pair of turning
points. On average, the classical particle spends equal amounts of time on
either side of the imaginary axis. Furthermore, the pattern of trajectories
changes qualitatively as the real part of the energy is varied. For some regions
of ${\rm Re}\,E$ the trajectories form a parity-symmetric pattern and these
energy regions interlace with other regions of ${\rm Re}\,E$ for which the
trajectories form a parity-anti-symmetric pattern.

Having observed the tunneling-like behavior of a classical particle with complex
energy in a double well, it is natural to ask what happens to a particle in a
periodic potential. An electron in a crystal lattice is described by such a
potential. An elementary physical system that also has a periodic potential
consists of a simple pendulum in a uniform gravitational field \cite{S16}.
Without loss of generality we take the pendulum bob to have mass $m=1$, the
string to have length $L=1$, and the uniform gravitational to have field
strength $g=1$. If the gravitational potential energy of the system is zero at
the height of the pivot point of the pendulum and if $x$ represents the angle
through which the pendulum bob swings, then the Hamiltonian $H$ for the pendulum
is
\begin{equation}
H=\half p^2-\cos x.
\label{e1}
\end{equation}
The classical equations of motion for this Hamiltonian are
\begin{equation}
\dot{x}=\frac{\partial H}{\partial p}=p,\qquad\dot{p}=-\frac{\partial H}{
\partial x}=-\sin x.
\label{e2}
\end{equation}

The Hamiltonian $H$ for this system is a constant of the motion, and thus the
energy $E$ is a time-independent quantity. If we take the energy $E$ to be real,
we find that the classical trajectories are confined to cells of horizontal
width $2\pi$. This is the periodic analog of the double-well anharmonic
oscillator. If the energy of the classical particle in a periodic potential is
taken to be complex, the particle hops from well to well in analogy to the
behavior of a quantum particle in a double well. This hopping behavior is a
deterministic random walk \cite{S11,S17}.

\begin{figure}
\includegraphics[scale=0.43, viewport=0 0 1024 668]{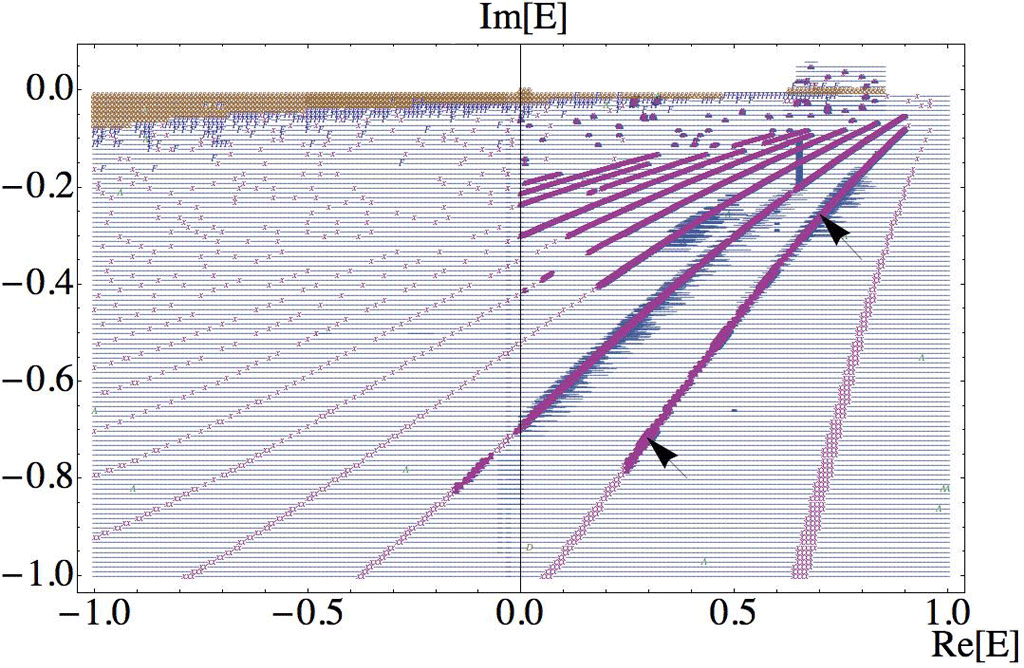}
\caption{Complex-energy plane showing those energies that lead to tunneling
(hopping) behavior and those energies that give rise to conduction for a
particle in a $-\cos x$ potential. Hopping behavior is indicated by a hyphen -
and conduction is indicated by an X. The symbol \& indicates that no tunneling
takes place; tunneling does not occur for energies whose imaginary part is close
to 0. In some regions of the energy plane very intensive numerical studies were
done and the X's and -'s are densely packed. This picture delineates the
features of band theory: If the imaginary part of the energy is taken to be
$-0.9$, then as the real part of the energy increases from $-1$ to $+1$, five
narrow conduction bands are encountered. These bands are located near ${\rm Re}
\,E=-0.95,\,-0.7,\,-0.25,\,0.15,\,0.7$. This picture is symmetric about ${\rm Im
}\,E=0$ and the bands get thicker as $|{\rm Im}\,E|$ increases. A total of 68689
points were classified to make this plot. In most places the resolution
(distance between points) is $0.01$, but in several regions the distance between
points is shortened to $0.001$. Note that the band edges are clearly and sharply
defined.}
\label{F1}
\end{figure}

The most dramatic analogy between complex classical mechanics and quantum
mechanics is established by showing that there exist narrow conduction bands in
the periodic potential for which the complex classical particle exhibits
unidirectional hopping and the quantum particle exhibits resonant tunneling.
The behavior of such a classical particle is discussed in detail in \cite{S17}.
In Ref.~\cite{S17} a lengthy numerical analysis was done that shows that the
classical conduction bands have a narrow but finite width (see Fig.~\ref{F1}).
The edges of the conduction bands are sharply defined. 

\section{Numerical Study of Elliptic Potentials}
\label{s2}

\begin{figure}
\begin{center}
\includegraphics[scale=0.47, viewport=0 0 1000 515]{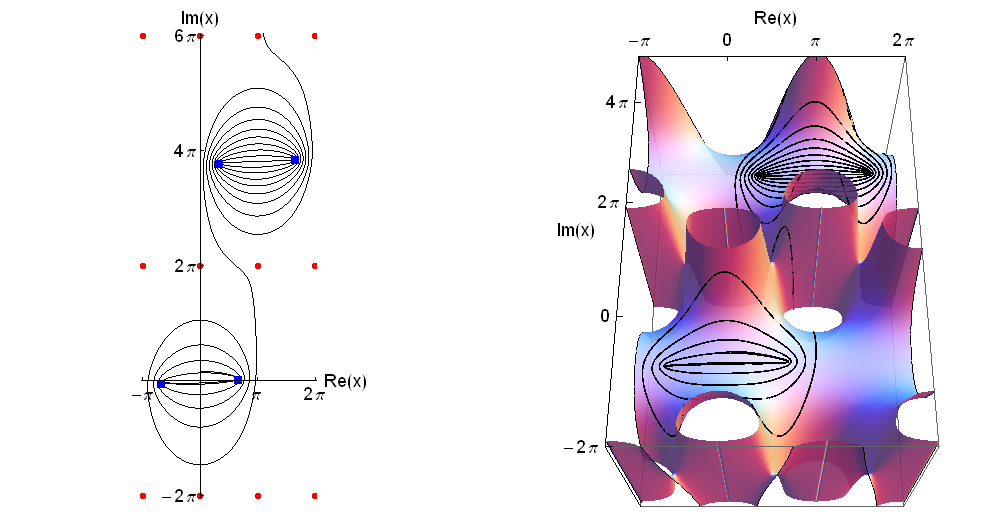}
\end{center}
\caption{Trajectory of a particle in the complex cnoidal potential ${\rm Cn}(x,
1/10\,000)$. The particle is initially at $x(0)=i$ and its path $x(t)$ is
plotted in the complex-$x$ plane for $0\leq t\leq 124.4$. The energy of the
particle is $E=1/2+i/10$. In the left panel the turning points are indicated by
small squares and the pole singularities of the cnoidal function are indicated
by small circles. In the right pane the particle trajectory is superimposed on a
three-dimensional relief plot of the real part of the cnoidal function.}
\label{F2}
\end{figure}

Having studied the behavior of a particle in a trigonometric periodic potential,
it is natural to generalize this analysis by considering the behavior of a
particle in an elliptic potential. Elliptic functions are doubly-periodic
generalizations of trigonometric functions. Also, they are analytic except for
simple poles, and this allows us to continue analytically the classical
trajectories into the complex plane. The specific Hamiltonian that we have
chosen to examine is the obvious generalization of the Hamiltonian in
(\ref{e1}):
\begin{equation}
H=\half p^2-{\rm Cn}(x,k).
\label{e3}
\end{equation}
Here ${\rm Cn}(x,k)$ is a {\it cnoidal} function \cite{S18,S19}. When the
parameter $k=0$, the cnoidal function reduces to the singly-periodic function
$\cos x$ and when $k=1$ the cnoidal function becomes ${\rm tanh}\,x$. When $0<
k<1$, the cnoidal function is periodic in both the real and imaginary directions
and it is meromorphic. (A {\it meromorphic} function is analytic in the finite-$
x$ plane except for well-separated pole singularities.) The real part of the
cnoidal potential ${\rm Cn}(x,k)$ is plotted in Fig.~\ref{F2}.

The classical particle trajectories satisfy Hamilton's equations
\begin{eqnarray}
\dot{x}&=&\frac{\partial H}{\partial p}=p,\nonumber\\
\dot{p}&=&-\frac{\partial H}{\partial x}=-{\rm Sn}(x,k){\rm Dn}(x,k).
\label{e4}
\end{eqnarray}
The trajectories for $k>0$ are elaborate in that the classical particles can
move vertically as well as horizontally. Figure \ref{F2} shows a typical
trajectory of a particle having complex energy. The energy of the particle is
$E=1/2+i/10$ and $k=1/10\,000$. We know from Ref.~\cite{S17} that if $k=0$, the
particle will spiral outward from a pair of turning points and then hop
horizontally to an adjacent site on the real-$x$ axis. It is remarkable that
even though $k$ is very small, the particle moves vertically. Thus, giving the
parameter $k$ a nonzero value represents a singular perturbation of the
Hamiltonian in (\ref{e1}). 

The right panel of Fig.~\ref{F2} shows the real part of the cnoidal function
${\rm Cn}(x,1/10\,000)$ plotted as a function of complex $x$. The poles of the 
cnoidal function appear as chimneys in both the positive and the negative
directions. Note that the path of the particle rolls around the singularities
until it is captured by a pair of turning points in another cell. The particle
spirals inward around these turning points and then spirals outward again before
leaving the cell and heading to another.

The novelty of the cnoidal potential is that complex classical-mechanical
particles may hop from cell to cell in both horizontal and vertical directions. 
While this motion appears to be two-dimensional, we emphasize that the
Hamiltonian in (\ref{e3}) has only one (complex) degree of freedom.

\begin{figure}
\begin{center}
\includegraphics[scale=0.33, viewport=0 0 1000 1260]{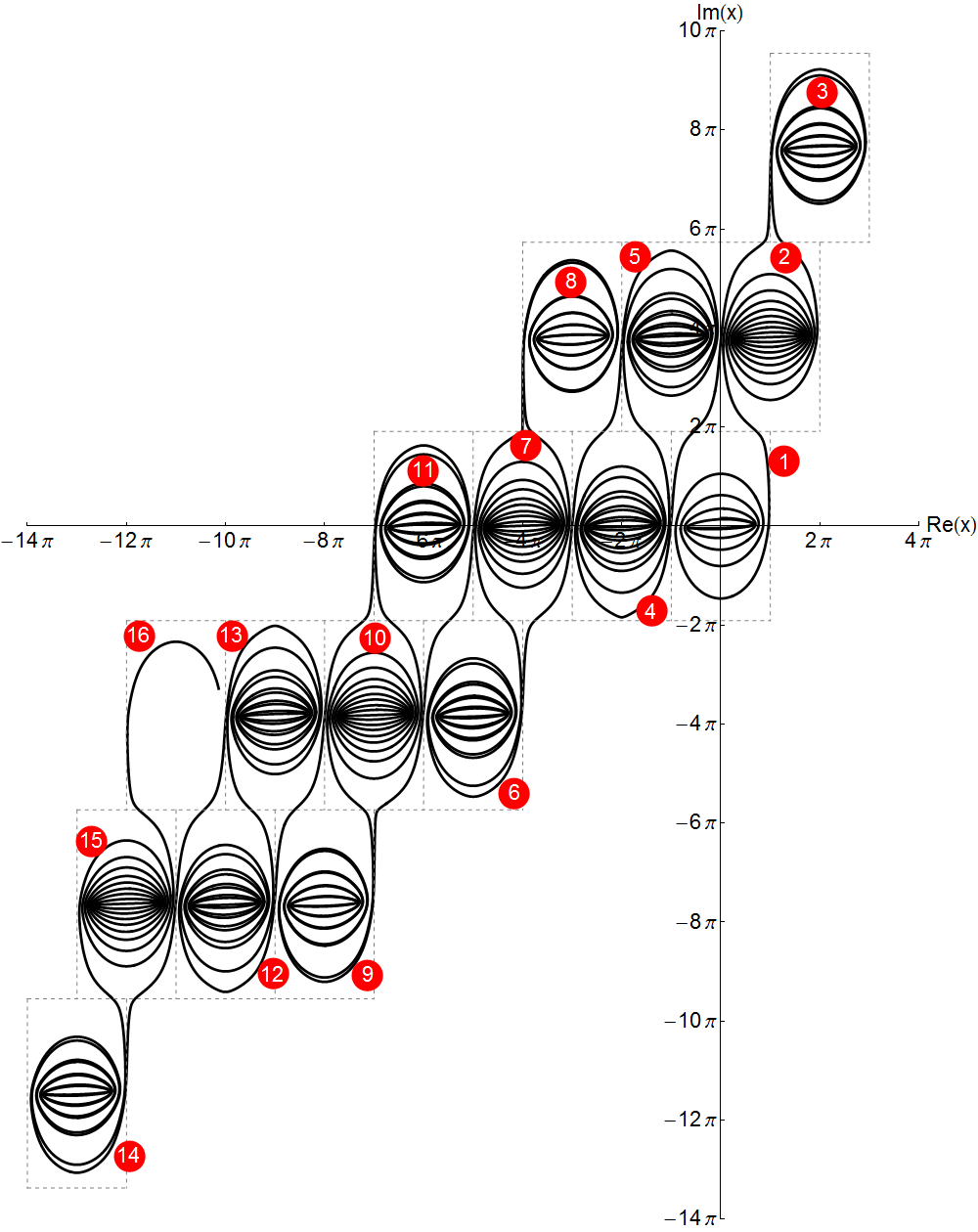}
\end{center}
\caption{Complex classical trajectory for a particle of energy $E=1/2+i/10$ in
a cnoidal potential with $k=1/10\,000$. The trajectory begins at $x(0)=i$ in
cell 1. (Cells are delineated by dotted lines and poles lie at the corners of
every cell. Each cell contains one pair of turning points.) The particle spirals
out of cell 1 and is captured by the turning points in cell 2. It then escapes
from these turning points and is captured by the turning points in cell 3, and
so on. Note that cell 1 is adjacent to cell 2, and that cell 2 is adjacent to
cell 3, but that cell 3 is {\it not} adjacent to cell 4. The path shown in this
figure reaches cell 16 at $t=1200$. The particle shown in this figure is never
captured by the same pair of turning points twice. This is because the energy of
the particle lies in a conduction band, as one can see from a careful
examination of Fig.~\ref{F1}.}
\label{F3}
\end{figure}

What happens to the trajectory of the particle in Fig.~\ref{F2} if we follow it
for a longer period of time? Figure \ref{F3} shows the trajectory for $0\leq t
\leq 1200$. Observe that the particle is temporarily captured by 16 cells in
this time period. These cells are numbered in the order in which their turning
points capture the particle, where by {\it capture} we mean that the particle
spirals around the turning points in that cell. After the particle leaves cell
1, it is temporarily captured by the turning points in cell 2, which is
northeast of and adjacent to cell 1. Similarly, cell 3 is northeast of and
adjacent to cell 2. However, the particle then jumps nonadjacently to cell 4,
passing briefly through cells 2, 5, and 1, without being captured by the turning
points in these cells.

Note that the particle in Fig.~\ref{F3} is never captured by the same pair of
turning points twice. The behavior of this complex self-avoiding trajectory is
typical of a particle whose energy lies in a conduction band. It is not easy to
verify it, but a careful examination of Fig.~\ref{F1} shows that the energy $E=
1/2+i/10$ lies in the middle of a conduction band. (Figure \ref{F1} was
constructed for $k=0$, but because $k=1/10\,000$ in Fig.~\ref{F3}, Fig.~\ref{F1}
gives an extremely good approximation to the location of the conduction bands.)

\begin{figure}
\begin{center}
\includegraphics[scale=0.44, viewport=0 0 1000 800]{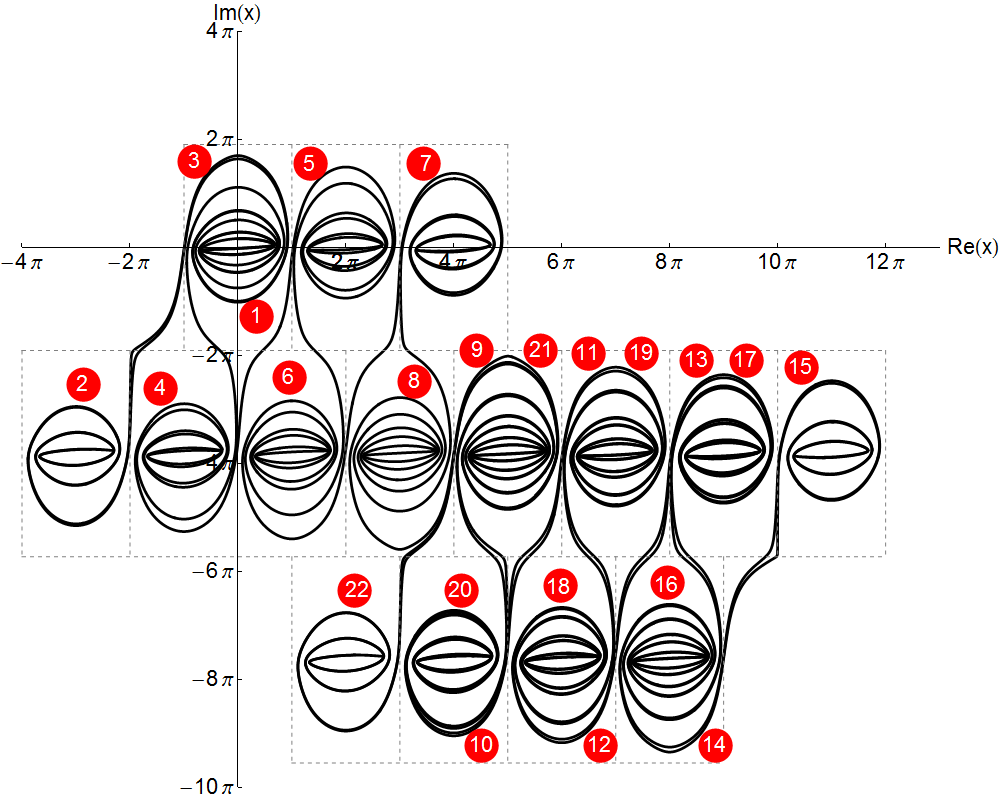}
\end{center}
\caption{Same as Fig.~\ref{F3} except that the energy of the particle is $E=1/2+
3i/20$. This energy does not lie in a conduction band, and consequently the
particle is captured twice by several pairs of turning points. The turning-point
captures are labeled sequentially by the numbers in the circles. As an example
of a double capture, when the particle is captured for the sixteenth time, it is
in the same cell as it was during its fourteenth capture. Note that complex
trajectory never crosses itself.}
\label{F4}
\end{figure}

To see how the trajectory of a particle behaves when the energy of the particle
is not in a conduction band, we change the energy of the particle in
Fig.~\ref{F3} to $E=1/2+3i/20$. This energy lies outside a conduction band, and
consequently the particle can be recaptured by pairs of turning points. In
particular, one can see in Fig.~\ref{F4} that as $t$ runs from $0$ to $1200$
the particle is captured twice by the turning points in six cells. This
particle appears to be executing a two-dimensional deterministic random walk.

\begin{figure}
\begin{center}
\includegraphics[scale=0.36, viewport=0 0 1000 685]{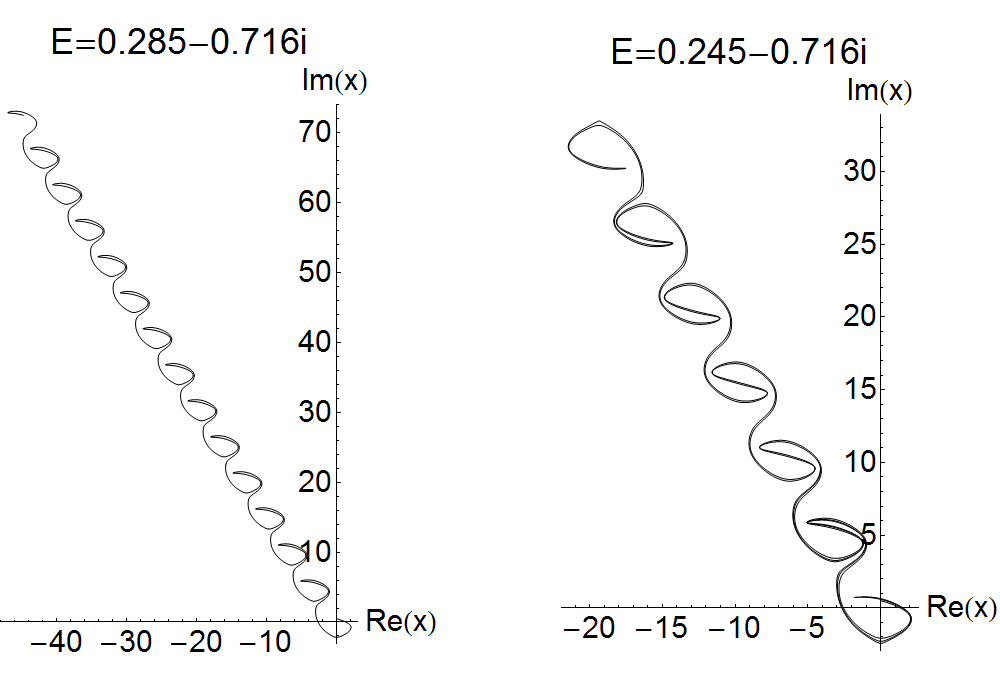}
\end{center}
\caption{Complex classical trajectories for particles of energy $E=0.285-0.716i$
(left pane) and $E=0.245-0.716i$ (right pane) in a cnoidal potential with $k=
1/10$. Both trajectories begin at $x(0)=i$ and run from $t=0$ until $t=200$. In
the left pane the energy is in a conduction band and the trajectory never
revisits any cell, but in the right pane the energy is not in a conduction band
and the trajectory is recaptured by previously visited turning points.}
\label{F5}
\end{figure}

To compare in greater detail the differences between conduction-band and
nonconduction-band trajectories, we investigate two trajectories that arise
from the initial point $x(0)=i$ when $k=1/10$. For the first trajectory we take
the energy to be $E=0.285-0.716i$ and for the second trajectory we take the
energy to be $E=0.245-0.716i$. An examination of Fig.~\ref{F1} shows that in the
former case the energy is in a conduction band and in the latter case the energy
is not. In Fig.~\ref{F5} we plot both trajectories for $0\leq t\leq 200$. The
first trajectory moves off in a northwesterly direction and never returns. In
the left pane of Fig.~\ref{F5} we show the first fourteen captures. The right
pane shows the second trajectory. This trajectory begins to revisit cells
starting with its seventh capture. We emphasize that the trajectory never
crosses itself. 

\begin{figure}
\begin{center}
\includegraphics[scale=0.32, viewport=0 0 1000 442]{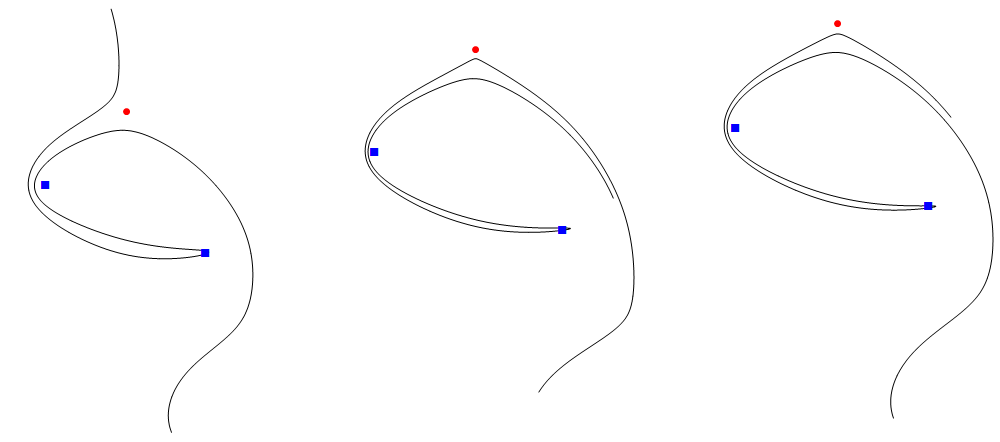}
\end{center}
\caption{Three classical trajectories similar to those in Fig.~\ref{F5}. All
three trajectories begin at the lower end of the figure. In the left pane the
trajectory loops around a pair of turning points (small squares) and is then
deflected by a pole singularity (small circle) upward into the next cell. The 
other two panes show the different ways for the trajectory to turn back on
itself.}
\label{F6}
\end{figure}

There are two possible ways for a trajectory to turn around when the energy of
the particle is not in a conduction band. These are shown in the middle and
right panes of Fig.~\ref{F6}. In all three panes in Fig.~\ref{F6} a trajectory
begins at the lower end of the figure and moves upward. In the left pane the
trajectory loops around the two turning points and encounters a pole
singularity, which deflects the trajectory upwards into the next cell. The next
two panes show trajectories that are deflected downward by a pole singularity;
these trajectories will be recaptured by a previous pair of turning points. The
trajectory in the middle pane differs from that in the right pane in that it
encircles the lower turning point in an anticlockwise direction and thus it
continues downward to the {\it left} of the original upward trajectory.

\begin{figure}
\begin{center}
\includegraphics[scale=0.33, viewport=0 0 1000 612]{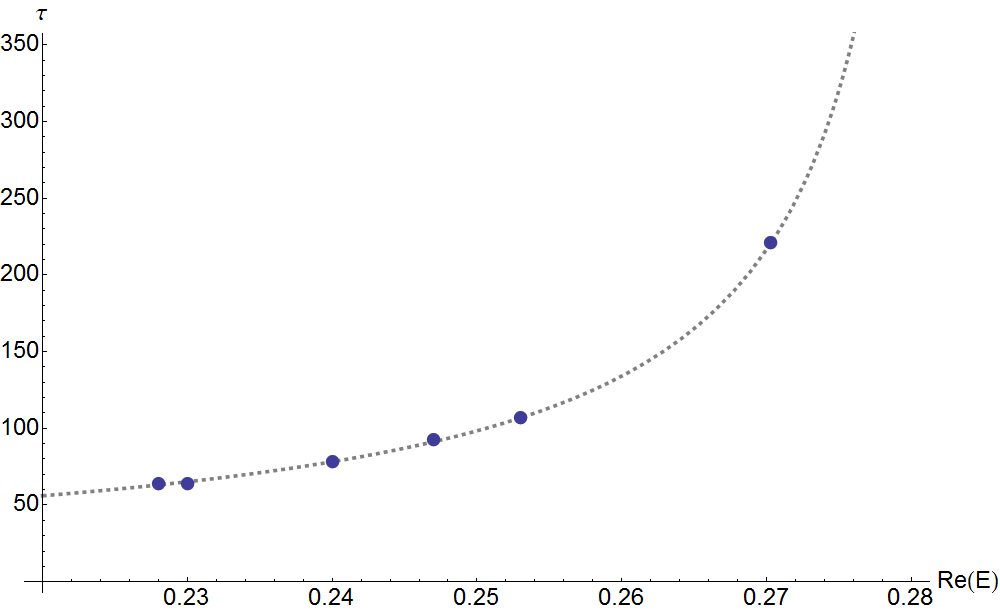}
\end{center}
\caption{Plot of the turn-around time $\tau$ for a complex trajectory near the
edge of a conduction band. The complex cnoidal potential has $k=1/10$ and $\tau$
is plotted as a function of the real part of the energy of the particle. The
imaginary part of the energy is held constant for all data points at ${\rm Im}\,
E=0.716$. The turn-around time becomes infinite at the edge of the conduction
band ${\rm Re}\,E=0.284$. The dotted line is the fitting curve in (\ref{e5}).}
\label{F7}
\end{figure}

To investigate what happens at the edge of a conduction band, we have done a
detailed numerical study of trajectories of particles having the same imaginary
energy component ${\rm Im}\,E=0.716$ but different real energies for a given
value of $k$ and $x(0)$. We have then determined the time required for a
particle trajectory to turn around as the real part of the energy nears the edge
of a conduction band. One would expect that it takes longer for the particle
trajectory to turn around, as it does in the right pane of Fig.~\ref{F5}, when
the real part of the energy approaches the band edge. Measuring the precise
turn-around time is difficult, and to do so we must look for turn-around
behavior of a consistent type; we have chosen to look for the behavior shown in
the {\it right} pane of Fig.~\ref{F6}. In Fig.~\ref{F7} we plot the turn-around
time $\tau$ as a function of the real part of the energy while holding the
imaginary part of the energy constant. The band edge is located at ${\rm Re}\,E=
0.284$. The dotted curve is a fit to the data points of the form
\begin{equation}
\tau({\rm Re}\,E)=\frac{4.872}{(0.284-{\rm Re}\,E)^{8/9}}.
\label{e5}
\end{equation}

\section{Concluding Remarks}
\label{s3}

There is a deep connection between quantum mechanics and complex classical
mechanics. In the complex domain the classical trajectories exhibit a remarkable
behavior that is analogous to quantum tunneling and periodic potentials exhibit
a surprising and intricate feature that closely resembles quantum band
structure. We have seen in this paper that the classical bands, just like the
quantum bands, have finite width and well-defined sharp edges.

The complex trajectories of a particle in a doubly-periodic potential can
exhibit a wider range of qualitative behaviors than those of a particle in a
singly-periodic potential partly because there is a two-dimensional array of
cells containing pairs of turning points and also because doubly-periodic
potentials have singularities. We have observed the following features: (i) The
path of a particle can go both vertically and horizontally as the particle
visits the cells in the complex-$x$ plane (see Fig.~\ref{F4}). (ii) When a
particle is captured by a pair of turning points in a cell, the number of turns
in the inward and outward spirals does not change from cell to cell, and thus
there is a characteristic capture-and-escape time (see
Figs.~\ref{F2}--\ref{F4}). This time depends on the energy of the particle and
on the value of $k$. (iii) If the energy of the particle does not lie in a
conduction band, the particle can be captured by pairs of turning points
multiple times (see Fig.~\ref{F4}). (iv) The particle trajectory is continuous,
and therefore each new cell that a particle visits is adjacent to the previous
cell. However, the particle is not always captured by the turning points in an
adjoining cell; captures may occur nonlocally (see Fig.~\ref{F3}). (v) Finally,
when the energy of a particle lies in a conduction band, the particle is never
captured twice by any pair of turning points. However, the trajectory of the
particle is not necessarily unidirectional; the particle may exhibit elaborate
kinds of zigzag motions as it drifts through the lattice (see Fig.~\ref{F3}).

A difficult and so far unanswered question is, Does a complex classical particle
whose energy is not in a conduction band undergo a two-dimensional random walk
in a doubly-periodic potential as it visits the cells in the lattice? The answer
to this question is not obvious because the trajectory of the particle cannot
cross itself. However, this does not mean that the particle must hop from cell
to cell in a self-avoiding manner. Numerical studies suggest that it is still
possible for the particle to visit cells repeatedly and in any order.

\vspace{0.5cm}
\footnotesize
\noindent
CMB is grateful to Imperial College for its hospitality and to the
U.S.~Department of Energy for financial support. DWH thanks Symplectic
Ltd.~for financial support. Mathematica was used to generate the figures in
this paper.

\end{document}